\documentstyle[aps,twocolumn,epsf]{revtex}

\catcode`\@=11

\newcommand{\fcaption}[1]{\vspace{1ex}
        \refstepcounter{figure}
        \setbox\@tempboxa = \hbox{\footnotesize {\bf ig.~\thefigure.} #1}
        \ifdim \wd\@tempboxa > 8cm
           {\begin{center}
        \parbox{8cm}{\footnotesize\baselineskip=8pt {\bf Fig.~\thefigure.} #1}

            \end{center}}
        \else
             {\begin{center}
             {\footnotesize {\bf Fig.~\thefigure.} #1}
              \end{center}}
        \fi}
\begin{document}
\title{Hubbard model on d-dimensional hypercubes: Exact solution 
for the two-electron case}

\vspace{2cm}

\author{Michel Caffarel$^{1}$ and R\'emy Mosseri$^{2}$}

\vspace{3cm}

\address{
$^{1}$ CNRS-Laboratoire de Chimie Th\'eorique
Tour 22-23, Universit\'{e} Pierre et Marie Curie
4 place Jussieu \\ 75252 Paris Cedex 05; France
e-mail: mc@lct.jusssieu.fr\\
$^{2}$ CNRS-Groupe de Physique des Solides
Universit\'{e} Paris 6 et Paris 7,
2 place Jussieu; 75251 Paris Cedex 05; France
e-mail: mosseri@gps.jussieu.fr}

\vspace{3cm}

\address{\rm (December 1997)}
\address{\mbox{ }}
\address{\parbox{14cm}{\rm \mbox{ }\mbox{ }
The Hubbard model is exactly solved for two particles with opposite spins
on d-dimensional hypercubes. It is shown that the spectrum can be separated 
into two parts: a trivial (U-independent) part resulting from symmetries of 
hypercubes and a non-trivial part expressed as a single-impurity 
problem on a set of finite chains of size $d^\prime+1$ ($d^\prime \le d$).
The exact expression for the one-particle 
Green's function is given. Finally, we discuss the extension of these 
results to standard hypercubic lattices with periodic boundary conditions.
}}
\address{\mbox{ }}
\address{\parbox{14cm}{\rm PACS No: 71.10+x ; 71.27+a}}
\maketitle

\makeatletter
\global\@specialpagefalse
\makeatother


One of the most widely used models to describe 
strongly correlated fermions systems is the Hubbard model 
and its various extensions. Unfortunately, 
only limited {\sl exact} information about its physical properties 
is available.  For the one-dimensional lattice the celebrated solution of Lieb 
and Wu \cite{Lieb} provides the exact eigenspectrum of the model.
In the opposite limit of large dimensions investigated very recently \cite{antoine}, 
some almost exact results have also been obtained.
However, for intermediate dimensions, and particularly for the very important 
two-dimensional case believed to be relevant to high-T$_c$ 
superconductivity, very little is at our disposal. Most of the results 
reported so far
have been obtained either from numerical solutions on very small clusters 
(subject to important finite-size effects)
or by using a variety of approximate analytical methods 
(with domains of validity and/or systematic errors difficult to evaluate)\cite{Dagotto}.

In this paper we present an exact solution for 
the Hubbard model with first-neighbor
hopping term $t$ and on-site interaction energy $U$
\begin{equation}
 H= -t \sum_{<ij> \sigma} c_{i\sigma}^+ c_{j\sigma}
 + U \sum_i c_{i\uparrow}^+ c_{i\uparrow} c_{i\downarrow}^+
            c_{i\downarrow}= T(t)+V(U),
\label{Hhub}
\end{equation}
for two particles with opposite spins on a d-dimensional hypercube $\gamma_d$.  
The hypercube $\gamma_d$ is defined as the set of $N=2^d$ sites
whose $d$ coordinates are either 0 or 1. Although our solution is yet 
limited to the two-electron case, to exhibit exact results for a truly 
interacting system in dimension greater than one is clearly 
of primary interest. This is particularly true
since hypercubes are related to usual cubic lattices 
with periodic boundary conditions: $\gamma_{2d}$ is topologically
equivalent to a d-dimensional hypercubic lattice
of linear size equal to 4 with periodic boundary conditions, noted $Z_4^d$
(i.e., $\gamma_2$
is equivalent to the 4-sites ring, $\gamma_4$ equivalent to the 2-dimensional 
4$\times$4 cluster with periodic conditions \cite{refR1,Angl}, etc..), while 
$\gamma_{2d+1}$ is equivalent to two translated linked copies of $Z_4^d$.
To solve the above Hamiltonian 
we first use an abelian subgroup of the full $\gamma_d$ point group, 
to block-diagonalize the initial matrix of size $2^{2d}$ into $2^d$ 
blocks of size $2^d$. Quite interestingly, these smaller submatrices 
correspond to a simple family of effective one-electron Hamiltonians 
defined on $\gamma_{d^\prime}$ ($d^\prime \le d$), some of which having a 
single-impurity site. In a second step, we show how to further reduce 
these matrices into smaller blocks of size $d^\prime +1$ (instead of $2^d$), 
corresponding to 
finite chains of size $d^\prime +1$ with one impurity site at one end and 
new specific hopping terms. 
Using a standard approach for impurity problems we provide a closed 
expression sum rule for the eigenspectrum. We also derive the exact expression 
for the one-particle Green's function. Finally, the extension of these 
results to hypercubic lattices with periodic boundary conditions is briefly
discussed. 

Let us define $\Lambda_d$, an abelian subgroup of the full  $\gamma_d$
point group, generated by the $d$ reflections into perpendicular 
$(d-1)-$dimensional planes meeting at the $\gamma_d$ centre,
\begin{equation}
\Lambda_d=
\{ \pi_i, i=1 \ldots d \;\;| \;\; {\pi_i}^2=E,
  \;\; \pi_i \pi_j= \pi_j \pi_i \;\; \} 
\label{eq2}
\end{equation}
where $E$ is the identity operator. It is easy to show that $\Lambda_d$
has $2^d$ elements (owing to the commutativity between the mirror
operations), and that the orbit of a generic point under $\Lambda_d$ is a
hypercube $\gamma_d$. Let us number the $\Lambda_d $ elements $L$ by
integers between 0 and $2^d-1$, in the following way: $L$ being a product
of mirrors $\pi_i$, let $L$ be the number, written in base-two, whose
corresponding $i^{th}$ digits are equal to one ($\pi_1$ is $L=1$,
$\pi_1 \pi_2 \pi_3$ is $L=7$, etc...). Let us also locate the sites according
to their numbering in base-two: The i$^{th}$ coordinate is equal to the 
i$^{th}$ digit.
Now, the action of the symmetry
operation $L$, onto a site $s$, noted as the ``special product'' 
$L \bullet s$, can be explicited: each
time a digit equals 1 in $L$, it switches the corresponding digit in $s$.
For example, in 3 dimensions, $7 \bullet 1=6$, with corresponds in
coordinates to  $\pi_1\pi_2\pi_3(0,0,1)=(1,1,0)$.
To construct the irreducible representations (irreps) and the character
table of $\Lambda_d$, it is useful to remark that this group is
isomorphic to $d$ times the tensorial product with itself of the
two-element group $Z_2=\{E,\pi_1\}=\Lambda_1$\cite{note1}. This is nothing else
than the group theory counterpart of the natural construction of $\gamma_d$
as two displaced copies of $\gamma_{d-1}$.
The $\Lambda_d$ irreps are 
easily obtained as tensor products of the two irreps $\Gamma_0$ and
$\Gamma_1$ of $Z_2$. As an illustration, the character table fo
$\Lambda_2= Z_2 \times Z_2$ is shown in Table \ref{character}. 
All the irreps are
uni-dimensional ($\Lambda_d$ is abelian), and the character table is
identical to standard Hadamard matrices\cite{Hada},
which are orthogonal matrices 
with elements $\pm1$ (note also that lines and columns of 
Hadamard matrices represent the non-interacting one-electron orbitals of the 
problem).
Note that the $N$
($=2^d$) irreps can also be labelled by integers $M$, through their
base-two decomposition, by ordering the different occurences of $Z_2$ in the
tensorial product, and fixing the corresponding digit to 0 or 1, whether
the irreps $\Gamma_0$ and $\Gamma_1$ occur.

\begin{table}[h]
\caption{Character table for $\Lambda_2$. The irreducible
representations are numbered by an integer M as defined in text.}
\begin{tabular}{rrrrrr}
\hline
\multicolumn{1}{c}{M}&
\multicolumn{1}{c}{}&
\multicolumn{1}{c}{E}&
\multicolumn{1}{c}{$\pi_1$}&
\multicolumn{1}{c}{$\pi_2$}&
\multicolumn{1}{c}{$\pi_1\pi_2$}\\
\hline
0   & $\Gamma_0 \times  \Gamma_0$ & 1 & 1 & 1 & 1 \\
1   & $\Gamma_1 \times  \Gamma_0$ & 1 &-1 & 1 &-1 \\
2   & $\Gamma_0 \times  \Gamma_1$ & 1 & 1 &-1 &-1 \\
3   & $\Gamma_1 \times  \Gamma_1$ & 1 &-1 &-1 & 1 \\
\hline
\end{tabular}
\label{character}
\end{table}
Let us consider first the one-electron tight-binding spectrum with
hopping term $t$. Each eigenstate belongs to a given irrep of $\Lambda_d$,
and reads 
\begin{equation}
 | M >= \frac{1}{\sqrt{2^d}} \sum_{R \in \Lambda_d} \chi^M_R 
 R( | 0 >)
   = \frac{1}{\sqrt{2^d}} \sum_{r=0}^{N-1} (-1)^{m_{M,r}} | r\bullet
 0>
\label{eq3}
\end{equation}
where $m_{M,r}$ is the number of digits equal to 1 simultaneously in
$M$ and $r$ (we have used the above defined special product $\bullet$
to denote the action of the symmetry operation $R$ onto the basis kets).
Since all the sites are equivalent, it is sufficient to consider the
interaction of the site number 0 with its $d$ first neighbors which
have exactly one digit equal to 1 and belong to the set
$V_d=\{v_j=2^j, j=0 \dots d-1 \}$.
As a result, one finds that the eigenenergies read
\begin{equation}
 E_M= \sum_{j=0}^{d-1} \chi^M_{v_j}= -t(d-2N_M) 
\label{eq4}
\end{equation}
where $N_M$ is the number of digits equal to 1 in $M$. Since $M$ runs
from 0 to $2^d-1$, the $\gamma_d$ spectrum consists of $d+1$ levels,
from $-dt$ to $dt$, with equal spacing $2t$. The degeneracy of a $E_M$
level is the standard binomial coefficient 
counting the number of sets of $N_M$ digits among $d$. 
As $d$ tends to infinity, one recovers a
gaussian-shaped spectrum, as for the infinite-dimensional cubic
lattice.

\begin{figure} \begin{center}
  \fbox{\epsfysize=5cm\epsfbox{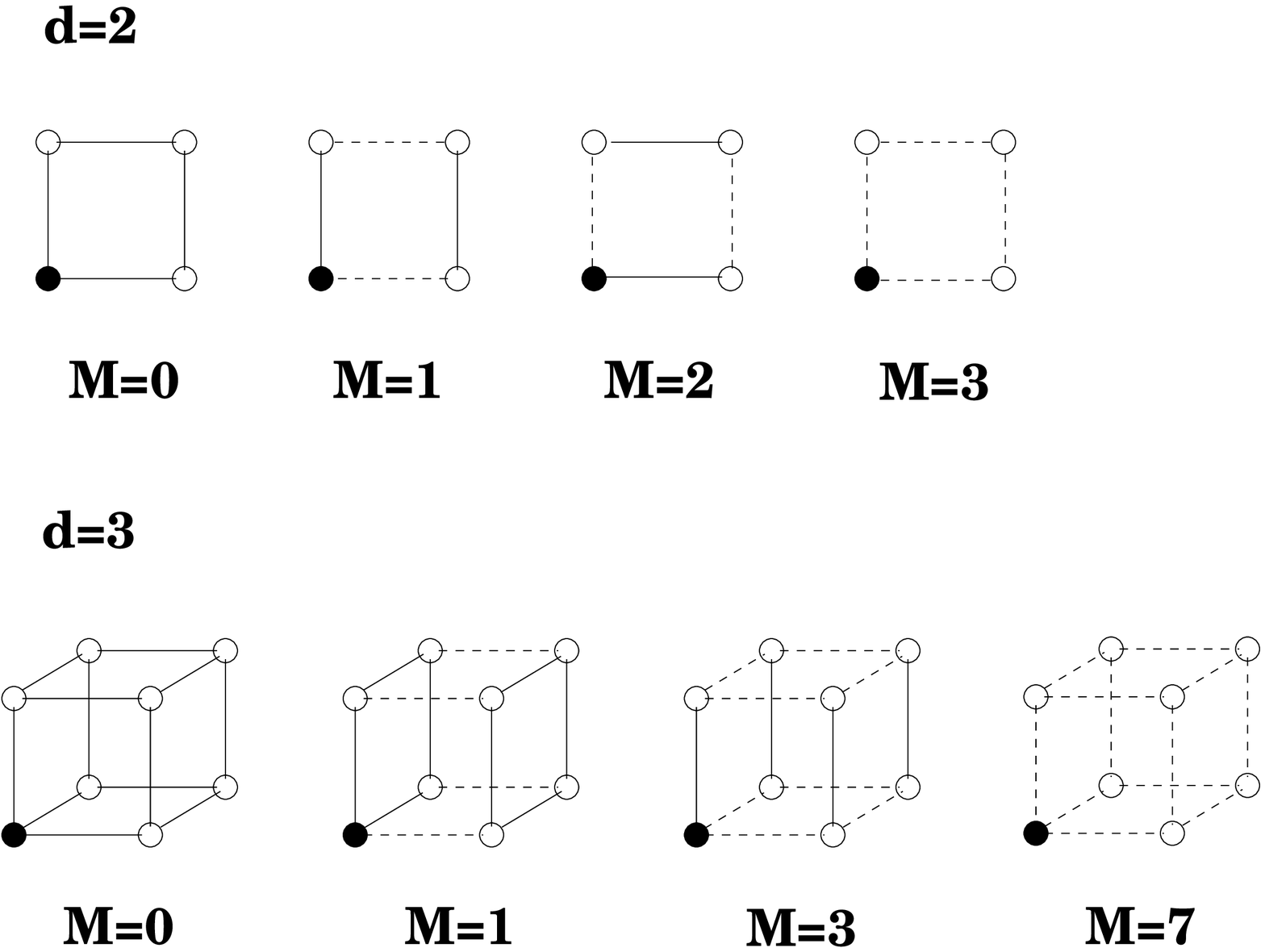}}
   \fcaption{
Graphical illustration of the effective Hamiltonian $H^{(M)}$
associated with the irrep $M$. Full circles correspond to
impurity sites with a diagonal term $U$, full edges to hopping terms
equal to $-2t$, and dotted lines to the edges which are cut in the
given irrep.

a) $d=2$, the four irreps, corresponding
to one square with one impurity, two times two segments, one containing
an impurity site, and finally four isolated sites (one being an impurity
with a diagonal term $U$).

b) the case $d=3$. Four irreps among the eight are shown.}

  \label{effHamilt}
\end{center}\end{figure}

Let us now consider the 
two-electron case (electrons with opposite spins). The full Hamiltonian
matrix, of size $2^{2d}$, is readily block-diagonalized into $2^d$ blocks
of size $2^d$, associated with each representation $M$. For a given $M$,
one constructs a basis with $2^d$ kets $| M, l>$
$$
| M, l >= \frac {1}{\sqrt{2^d}} \sum_{R \in \Lambda_d}
       \chi_R^M R( | 0\uparrow, l\downarrow >)
$$
\begin{equation}
       =\frac{1}{\sqrt{2^d}} \sum_{r=0}^{N-1}
       \chi_r^M | (r\bullet 0)\uparrow, (r \bullet l)\downarrow >
\label{eq5}
\end{equation}
It is clear that $U$ only occurs in $ < M,0 | V(U) | M,0 > $,
while the kinetic part reads
$$
<M,l | T | M, l^\prime >=  
$$
$$
\frac{1}{2^d} \sum_{r=0}^{N-1}
\sum_{r^\prime=0}^{N-1} \chi_r^M \chi_{r^\prime}^M
<(r\bullet 0)\uparrow(r \bullet l)\downarrow | T |
(r^\prime \bullet 0)\uparrow, (r^\prime \bullet l^\prime)\downarrow >
$$
$$
=  \frac{1}{2^d} \sum_{r=0}^{N-1} [ \chi_r^M \chi_{r}^M
 < (r \bullet 0) \uparrow (r \bullet l)\downarrow | T |
(r \bullet 0)\uparrow, (r \bullet l^\prime)\downarrow > 
$$
$$
+
\sum_{j=0}^{d-1}  \chi_r^M \chi_{v_j \bullet r}^M
$$
\begin{equation}
<(r\bullet 0)\uparrow(r \bullet l)\downarrow | T |
(v_j \bullet r \bullet 0)\uparrow, (v_j \bullet r \bullet l^\prime)\downarrow >
]
\label{eq6}
\end{equation}
In the latter expression, we have decoupled the terms where the $\downarrow$
spin jumps (first part) from those where the $\uparrow$ spin jumps. In the
latter part (sum over $j$) the only nonvanishing term corresponds to
$r \bullet l = v_j \bullet r \bullet l^\prime \rightarrow
 l^\prime= v_j \bullet l $, which means  $l$ and $l^\prime$ neighbors
(recall that the group is abelian and its elements are their own
inverse).
One finally gets that the nonvanishing kinetic terms read
\begin{equation}
    <M,l | T | M,v_j \bullet l >= -\frac{t}{2^d}
    \sum_{r=0}^{N-1} (1 + \chi_r^M \chi_{v_j \bullet r}^M )
\label{eq7}
\end{equation}
The block matrix associated to the irrep $M$ corresponds therefore
to an effective one-electron tight-binding Hamiltonian, with sites
labelled by $l$, kets $| M,l>$ and hopping terms given by the above
expression (\ref{eq7}).
A simple case is provided by the identity irrep $M=0$, for which
$ \chi_r^M =1$ for any $r$. The effective Hamiltonian $H^{(0)}$
corresponds to a  hypercube $\gamma_d$, with one ``impurity'' site
(with diagonal term $U$) and a constant hopping term $-2t$.
We now show that the other irreps correspond to effective Hamiltonian
$H^{(M)}$ on hypercubes $\gamma_{d^\prime}$ with $d^\prime<d$. One has to
evaluate the right-hand part of equation (\ref{eq7}). 
By definition of the $v_j$
(mirrors which connect a site to one of its first neighbors), $r$ and
$v_j \bullet r$ differ by one digit, which depends on $j$, not on $r$.
If this digit takes the
value 1 in the base-two decomposition of $M$, then
$\chi_r^M \chi_{v_j \bullet r}^M=-1$ for any $r$,
and the corresponding $H^{(M)}$ matrix element vanishes. $H^{(M)}$
therefore corresponds to a $\gamma_d$ in which families of parallel
edges have been cut (those orthogonal to the mirrors labelled by the
digits 1 in the decomposition of $M$). Figure \ref{effHamilt} displays graphically
this general result for $d=2$ and $d=3$.

So, at this step, the spectrum is solved in terms of one-electron 
tight-binding models on hypercubes of dimension lower or equal to $d$. The
only non-trivial part corresponds to those part, in the hypercube
decomposition, which contain the impurity site\cite{note2}.
We now show how to
further greatly reduce their complexity.

To do that it is convenient to re-express the effective Hamiltonian $H$
(hereafter  we will drop the superscript $(0)$ since only the $M=0$ case has 
finally to be considered) in the one-electron basis, Eq.(\ref{eq3}).
It is easy to verify that $H$ can be written in the form
\begin{equation}
H=  Diag({\epsilon}) + \frac{U}{N}  {I}_{N}  
\label{eq8}
\end{equation}
where $Diag(\epsilon)$
is a diagonal matrix whose entries are the $2^d$ free electron-solutions 
of the hypercube with hopping $2t$ (solution at $U=0$) and 
$I_N$ is the $N \times N$ matrix with unit entries for all $i$ and $j$.
The latter part is nothing but a projection 
operator on the eigenstate $|v_0>$ with energy $U$ and $H$ can be rewritten 
\begin{equation}
H=  Diag({\epsilon}) + U |v_0><v_0| 
\label{eq9}
\end{equation}
with $(v_0)_i = \frac{1}{\sqrt{N}}\;\; i=1, \ldots ,N.$

\begin{figure} \begin{center}
  \fbox{\epsfxsize=8cm\epsfbox{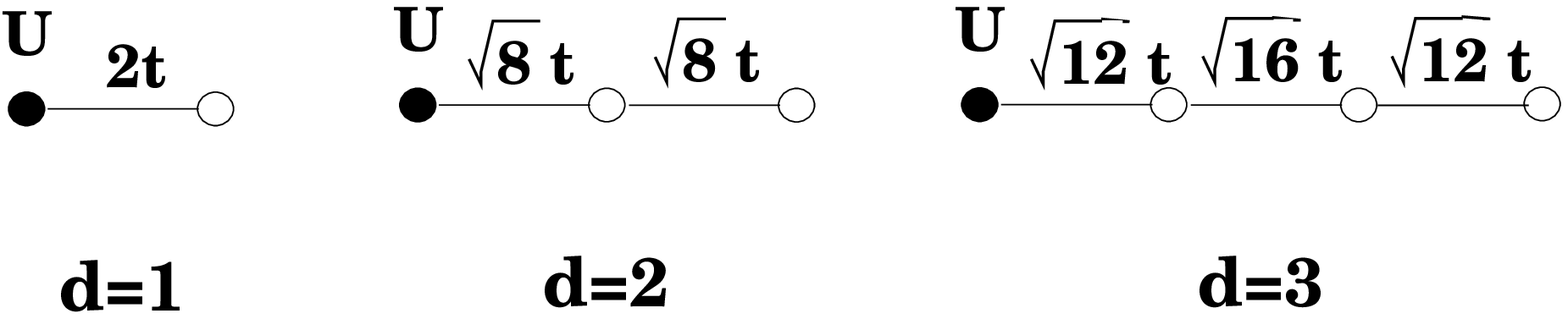}}
   \fcaption{
Graphical illustration of the single-impurity one-electron problems 
associated with the non-trivial part of the spectrum (general formula given
in text).}
  \label{eff2}
\end{center}\end{figure}

To proceed further we define for each subspace corresponding to a given 
value $\epsilon_i$ ($d+1$ subspaces of degeneracy $g_i$, the binomial 
coefficient) a basis consisting of the normalized vectors 
$v_1^{(i)}=\frac{1}{\sqrt{g_i}}(1, \ldots ,1) $ and a set 
of vectors \{$v_k^{(i)}, k\ne 1$\} spanning the subspace orthogonal to 
$v_1^{(i)}$. In this new basis 
only the ($d+1$) vectors $v_1$ have a non-zero overlap with $|v_0>$, and 
$H$, of dimension $2^d$, decomposes into a diagonal matrix 
having $2^d-(d+1)$ trivial 
solutions $\epsilon_i$, and a residual U-dependent part, noted $\cal{H}$, 
given by the matrix of linear size ($d+1$) written in the form
\begin{equation}
{\cal{H}}=  Diag({\epsilon}) +  U |v><v|  \equiv {\cal{H}}_0 + {\cal{V}}
\label{eq10}
\end{equation}
where $|v>$ has components $v_i=\sqrt{g_i/N}$ and the ${\epsilon}$ 's
represent now the (d+1) {\sl distinct} free-electron energies:
\begin{equation}
\epsilon_i = -2t(d-2i) \;\;\; i=0, \ldots ,d.
\label{eq11}
\end{equation}
An alternative representation consists in going back to the basis where 
the potential operator $\cal{V}$ is diagonal. $H$ is then found to 
be tridiagonal 
(one-dimensional tight-binding model) with off-diagonal hopping
term given by $t_{i,i+1}= 2t\sqrt{(d-i)(i+1)},\;\;i=0\;to\;d-1$ and a 
diagonal $U$-contribution at the initial site (here numbered 0). In other words,
the problem is mapped onto a single-impurity problem on a finite chain of 
linear size ($d+1$). Figure \ref{eff2} illustrates this result for $d \le 3$.
It is known that eigenspectrum of single-impurity problems can be expressed 
under the form of a sum rule expression using the Koster-Slater approach 
\cite{Koster}. This result can be briefly re-obtained as follows. First, the 
following operator identity \cite{note3} is invoked
$$
{({\cal{H}}-z)}^{-1}= {({\cal{H}}_0-z)}^{-1} 
$$
\begin{equation}
- \frac{U}{1 + U G_{vv}^{(0)}(z)} \frac{1}{({\cal{H}}_0-z)} |v><v|
\frac{1}{({\cal{H}}_0-z)} 
\label{eq12a}
\end{equation}
where $z$ is an arbitrary complex number and $G_{vv}^{(0)}$
the unperturbed  
Green's function given by $G_{vv}^{(0)}(z)= <v| {({\cal{H}}_0-z)}^{-1} |v>$.
Then, using this identity the fully interacting Green's function,${G}_{vv}(z)$, 
can be written as
\begin{equation}
{G}_{vv}(z)= \frac{{G}_{vv}^{(0)}(z)}{1 + U {G}_{vv}^{(0)}(z)}
\label{eq15}
\end{equation}
Searching for poles of ${G}_{vv}(z)$ we get
\begin{equation}
\frac{1}{N} \sum_{i=0}^{d} \frac{g_i}{E-\epsilon_i} = \frac{1}{U} \;\;.
\label{eq16}
\end{equation}
Eq.(\ref{eq16}) is the final closed expression determining the non-trivial 
part of the spectrum. It is easily shown that this equation admits ($d+1$) 
distinct solutions we shall denote 
as $Spect^{(d)}(U)$.

To summarize, the set of $2^{2d}$ eigenvalues of the hypercube 
$\gamma_d$ consists of a U-dependent 
part given by the collection of the non-trivial spectra at dimensions lower 
or equal to $d$:
\begin{equation}
\bigoplus_{i=0}^{d} \left( \begin{array}{c} d \\ i \end{array} \right) 
Spect^{(i)}(U)
\label{eq16a}
\end{equation}
(note that here the binomial coefficient counts the number of 
occurences of $\gamma_i$ 
hypercubes with an impurity site in the decomposition of the initial 
$\gamma_d$, see Fig.\ref{effHamilt})
and a trivial U-independent part given by:
\begin{equation}
E_l=2tl \;\;\;\;\;\;\;\;\;\;\;\;\; l=-d \ldots d
\end{equation}
with degeneracy:
\begin{equation}
g_l= {{\sum}^\prime}_{i=|l|}^{d} \left[ 2^{d-i} 
\left( \begin{array}{c} i \\ \frac{l+i}{2}  \end{array} \right) -1 \right]
\left( \begin{array}{c} d \\ i \end{array} \right) 
\label{eq16b}
\end{equation}
the prime in $\sum^{\prime}$ indicating that the summation over $i$ is done with 
an increment of two.
Formula (\ref{eq16b}) can be obtained by tracing back the 
contributions due to the different irreps 
and those issued from the internal symmetry of hypercubes
(trivial solutions extracted when passing from Eq.(\ref{eq9})) to 
Eq.(\ref{eq10})). The ground-state energy is not degenerate and 
corresponds to the lowest solution in $Spect^{(d)}(U)$. 
Note also that in the limit of large dimensions,
and after proper renormalization of the 
parameters of the Hamiltonian, an infinite-dimensional model can be 
defined, as done recently for a number of 
correlated fermions models, see \cite{antoine}.

Let us now consider dynamical properties. We are 
interested in evaluating the one-particle Green's function  defined by
\begin{equation}
G^{(d)}_k(z,U) \equiv < \psi_0|a_k \frac{1}{H-z} a^+_k|\psi_0>
\label{eq17}
\end{equation}
where $|\psi_0>$ denotes the one-particle ground-state consisting of 
one electron of given spin and $a^+_k$ creates 
one electron of opposite spin in a one-particle state, noted $k$. 
Hereafter, $k$ varies from 0 to $d$  and labels one of the 
($d+1$) degenerate subspaces 
of the one-electron problem, the ground-state corresponding to $k=0$.
Now, remark that 
vector $a^+_k |\psi_0>$ belongs to the subspace corresponding to the 
decomposition on hypercubes of dimension ($d-k$). 
There exist $2^k$ different families of 
hypercubes having such a dimension (See, fig. \ref{effHamilt}) 
among which only one has the 
U-impurity site. Accordingly, we get the following result 
\begin{equation}
G^{(d)}_k(z,U) = \frac{1}{2^k} G^{(d-k)}_0(z,U) + (1-\frac{1}{2^k})
 G^{(d-k)}_0(z,U=0).
\label{eq18}
\end{equation}
Now, we need to evaluate the fundamental quantity
$G^{(d^\prime)}_0(z,U)$ with Hamiltonian (\ref{eq10}), the dimension 
$d^\prime$ ranging from $0$ to $d$. For that we note
that the ket $|0>\equiv a^+_0 |\psi_0>$ represents the (first) 
basis element corresponding 
to the diagonal energy $\epsilon_0$ in representation (\ref{eq10}).
Projecting out identity (\ref{eq12a}) onto vector $|0>$ and 
expressing the different quantities in terms of the non-interacting 
spectrum we get:
$$
G^{(d^\prime)}_0(z,U)=
$$
\begin{equation}
\frac{1}{ {\epsilon}^{(d^\prime)}_{0} -z }
-\frac{U}{ 2^{d^\prime} {( {\epsilon}^{(d^\prime)}_{0} -z )}^2
( 1 + 
\frac{U}{2^{d^\prime}} \sum_{i=0}^{d^\prime} 
\frac{g_i}{{\epsilon_i}^{(d^\prime)} -z }) }
\label{eq20}
\end{equation}
This equation together with Eqs.(\ref{eq11}) and (\ref{eq18}) 
give the exact one-particle Green's function of the problem.

Generalization of this approach to higher fillings
is presently under investigation. We have already found that
the existence of large fractions of U-independent eigenvalues is still
valid for some specific fillings. However, the underlying
structure for the U-dependent part of the spectrum is more
difficult to elucidate.

Back to the two-electron case, we would like to mention that, by using 
rotations instead of reflections, a similar calculation can be done 
for standard d-dimensional hypercubic lattices  
of linear size L and periodic boundary conditions ($Z_L^d$). 
We have found that the problem can still be mapped onto a family of $L^d$ 
single-impurity one-electron effective problems defined on 
the very same lattice, $Z_L^d$.  
Each irrep corresponds to a value of a d-dimensional integer vector $\vec{M}$
$(M_i=0 \ldots L-1)$, 
or more physically, to a value for the total momentum $\vec{K}=\frac{2\pi}{L}\vec{M}$. 
The only difference between irreps lies in the value of the hopping term which 
depends explicitly on $\vec{M}$. More 
precisely, for each irrep labelled by $\vec{K}$, Eq.(\ref{eq16}) now reads
\begin{equation}
\frac{1}{L^d} \sum_{n_1=0}^{L-1}\ldots \sum_{n_d=0}^{L-1}
\frac{1}{E -\epsilon^{(\vec{K})}_{\vec{n}}} = \frac{1}{U} 
\label{eq21}
\end{equation}
where the non-interacting energies $\epsilon^{(\vec{K})}$ are given by
\begin{equation}
\epsilon^{(\vec{K})}_{\vec{n}} = -4t \sum_{i=1}^{d} \cos{(K_i/2)} \cos{(k_i 
+ K_i/2)}
\label{eq22}
\end{equation}
with $\vec{K}=\frac{2\pi}{L}\vec{M}$ and $\vec{k}=\frac{2\pi}{L}\vec{n}$. 
In contrast with hypercubes, the sum in 
Eq.(\ref{eq21}) runs over $L^d$ values and cannot be further reduced. 
Note that, since the effective hopping term varies with the irrep (and 
can even vanish), both localized and resonant states (irreps by irreps) may 
be present.
Finally, it is worth noticing that in the $d=1$ case the Bethe {\sl ansatz} 
equations for two particles\cite{Lieb}:
\begin{equation}
\tan{\frac{k_1 L}{2}} = \frac{U}{2t (\sin{k_1}-\sin{k_2})}
\end{equation}
with $E=-2t(\cos{k_1}+\cos{k_2})$ and $k_1+k_2 = \frac{2\pi}{L}M$ ($M=0 \ldots L-1$) 
can be recovered from Eq.(\ref{eq21}) after simple 
but tedious algebra. 

This work was supported by the ``Centre National de la Recherche Scientifique''
(C.N.R.S.)


\begin{references}
\bibitem{note1} For $d=4$,  $\Lambda_4$ coincides with the group $\cal{A}$ 
used in ref.\cite{Angl} (Appendix 1) where the full $\gamma_{4}$ symmetry 
group is used to diagonalize the Hamiltonian.
\bibitem{note2} The occurence of a large fraction of trivial states was 
noted for $d=2$, \cite{Angl}. Our results put new light on this topic 
for general $d$.
\bibitem{note3} Identity  obtained by making a series 
expansion of operator ${({\cal{H}}-z)}^{-1}$ in terms of ${\cal{V}}$ and 
using the fact that ${\cal{V}}$ is a projector.
\end{references}
\end{document}